\documentclass[12pt]{article}
\usepackage[utf8]{inputenc}
\usepackage{graphicx}
\usepackage{geometry}
\usepackage{amsmath}
\usepackage{footnote}
\usepackage{mathtools}
\usepackage{booktabs}
\usepackage{fancyhdr}

\usepackage[font=small,labelfont=bf]{caption}
\usepackage{tikz}
\usepackage{framed}
\usepackage{amsmath}
\usepackage{autobreak}
\usepackage{algorithm}
\usepackage[noend]{algpseudocode}
\usepackage{indentfirst}
\usepackage{tikz}
\usetikzlibrary{quantikz}
\usepackage{import}
\usepackage[
backend=biber,
style=alphabetic,
sorting=ynt
]{biblatex}
\AtBeginBibliography{\small}
\usepackage{xr-hyper}
\usepackage{hyperref}
\usepackage{subfiles}
\begin{document}

\hypersetup{
    colorlinks=true,
    linkcolor=[rgb]{0,0,0},
    filecolor=magenta,      
    urlcolor=cyan,
    pdftitle={Depth-First Grover Search Algorithm on Hybrid Quantum-Classical Computer},
    pdfpagemode=FullScreen,
    }

\begin{titlepage}
\centering
\vspace*{\fill}
\vspace*{0.5cm}

\Large\bfseries
   Depth-First Grover Search Algorithm \\ on Hybrid Quantum-Classical Computer
    
\vspace*{0.5cm}

\large Haoxiang Guo\\
yurodidon@gmail.com\\

\vspace*{\fill}
\end{titlepage}

\newpage
\pagenumbering{roman}

\pagestyle{fancy}
\fancyhf{}
\lfoot{Haoxiang Guo 2022}
\rfoot{Page \thepage}

%Abstract
\begin{abstract}
    We demonstrated the detailed construction of the hybrid quantum-classical computer. Based on this architecture, the useful concept of amplitude interception is illustrated. It is then embedded into a combination of Depth-First Search and Grover's algorithm to generate a novel approach, the Depth-First Grover Search(DFGS), to handle multi-solution searching problems on unstructured databases with an unknown number of solutions. Our new algorithm attains an average complexity of $\mathcal{O}(m\sqrt{N})$ which performs as efficient as a normal Grover Search, and a $\mathcal{O}(\sqrt{p}N)$ complexity with a manually determined constant $p$ for the case with all elements are solutions, where a normal Grover Search will degenerate to $\mathcal{O}(N\sqrt{N})$. The DFGS algorithm is more robust and stable in comparison.
\end{abstract}
%Keywords
\begin{sloppypar}
\section{Introduction}
{
    \par Since the concept of Quantum Computing was proposed\cite{QuantumComputation,Feynman1985}, bunches of quantum algorithms were designed to achieve quantum supremacy\cite{QS}. And these algorithms can be separated into two large groups: \textbf{(1)} Algorithms based on Quantum Fourier Transformation\cite{QFT, Shor, Shor1997} and \textbf{(2)} Algorithms based on oracles\cite{DJ1992, Bernstein1997, Grover1996}, in which Grover's algorithm is able to solve the single-solution unstructured searching problems with a complexity of $\mathcal{O}(\sqrt{N})$, and was demonstrated to be asymptotic optimal in \cite{Bennett1997}.
    \par In recent years, a new concept of Hybrid Quantum-Classical(HQC) Computation was presented\cite{HQC} and received rising degrees of attention, the concept of HQC is applicated to several areas in computer science\cite{hcq1, hcq2, hcq3, hcq4}. By attaching quantum components with a classical computer, two parts complement each other, leading HQC to have both merits of them, e.g.,  quantum parallelism\cite{ni2010-fs}, data storing, and efficient arithmetic operations.
    \par Whereas a few articles talk about the detailed structure of HQC, in this paper, we devoted Section 2 to investigating the configuration of HQC. Furthermore, we confronted the issue of the inefficiency of Grover's algorithm when applying it to multi-solution search problems (which will encounter repetitions and worsen up to $\mathcal{O}(N\sqrt{N})$), and proposed the concept of amplitude interception and a novel robust algorithm, the Depth-first Grover Search based on an HQC computer in Section 3.
}

\section{Hybrid Quantum-Classical Structure}
{
    \par We aim to design an architecture for HQC with two basic functions:
    \begin{itemize}
        \item The ability to operate quantum parts and classical parts individually.
        \item The ability to intercommunicate between the quantum part and the classical part.
    \end{itemize}
    \par Both of these functions can be implemented by two components: $\mathtt{INITIALIZER}$ and $\mathtt{ENCODER}$. $\mathtt{INITIALIZER}$ component initializes auxiliary qubits for follow-up quantum computing, e.g., for handling searching problems, it is usually composed with $HX$ gates to get $\ket{-}$ for phase kickback\cite{CEMM1998}. Besides the $\mathtt{INITIALIZER}$, the $\mathtt{ENCODER}$ component enables the information transformation from classical to quantum parts. An example is shown in Figure 1, which directly copies the digits to qubits.
    \begin{figure}[htbp!]
        \centering
        \begin{quantikz}
            \lstick{\ket{0}} & \gate{X} & \ \ldots\ \qw & \qw &
            \qw \rstick{\ket{q_1}}\\
            &&\ldots\\
            \lstick{\ket{0}} & \qw & \ \ldots\ \qw & \gate{X} &
            \qw \rstick{\ket{q_n}}\\
            \lstick{$c_1$} & \cw \cwbend{-3} & \ \ldots\ \cw & \cw &
            \cw & \cw \\
            &&\ldots\\
            \lstick{$c_n$} & \cw & \ \ldots\ \cw & \cw \cwbend{-3} &
            \cw & \cw \\
        \end{quantikz}
        \caption{An example of $\mathtt{ENCODER}$, composed with $n$ control-not gate}
    \end{figure}
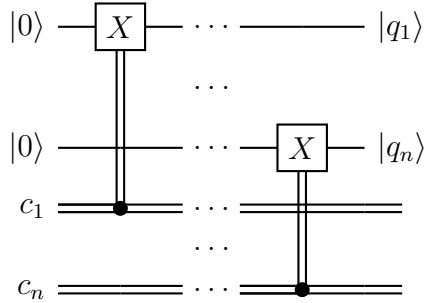
    \par The bridge from the classical part to the quantum part is also proven to have various forms. In \cite{suchara2018hybrid}, the quantum part is considered to be arranged in the cloud and intercommunicated with Internet technology. Recently, IBM constructed 127-qubits quantum computers with remote accessibility\cite{ibmqc}, which implies that future quantum computing will have a trend of online-offline separation. The general architecture of HQC is demonstrated in Figure 2.
    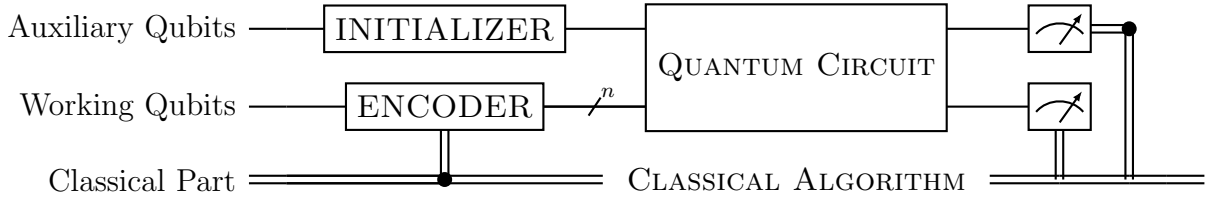
\begin{figure}[h]
        \begin{center}
            \begin{quantikz}
                \lstick{Auxiliary Qubits} & \qw & \gate{\textsc{INITIALIZER}} & \gate[wires=2][4cm]{\textsc{Quantum Circuit}} & \meter{} & \cwbend{2} \\[-0.2cm]
                \lstick{Working Qubits} & \qw & \gate{\textsc{ENCODER}} & \qwbundle{n}  & \meter{} \vcw {1} \\
                \lstick{Classical Part} \cw & \cw & \cw \cwbend{-1} & \cw \text{\textsc{\ \ Classical Algorithm\ \ }} & \cw & \cw & \cw & \cw 
            \end{quantikz}
        \end{center}
        \vspace{-0.2cm}
        \caption{The detailed structure of the hybrid quantum-classical computer. In many cases, the auxiliary qubits are not necessary to be measured.}
    \end{figure}
    \par A crucial feature of this architecture is it allows the implementation of discrete loops\cite{Blieberger1994}, we refer to discrete since we always destroy superpositions after each measurement. Since an amount of information is inevitably lost during the measurements, so there exists an irremovable gap between the conditional loops we commonly regard and the discrete loops we have. Or in other words, we are still unable to implement conditional loops upon this architecture. 
    \par In HQC, classical parts can either work as the main parts for performing algorithms or auxiliary parts for data storing and updating. A sort of classical algorithm can be improved by replacing some steps(e.g., extreme searching\cite{DH96, AAKS, KOWADA2008}) with quantum algorithms to obtain better performance\cite{Hei03, AS05}. Furthermore, some other authors proposed variational hybrid quantum-classical algorithms\cite{McClean2016, Plekhanov2022} by regarding classical computers as an auxiliary part. As well as in \cite{Bravyi2016} the authors discovered a portion of qubits required in quantum algorithms could be separated and replaced by classical bits with the same efficiency but save quantum resources.
}

\section{Improvement of Grover's Algorithm}
{
    {
    \par We aim to design a novel algorithm, with classical and quantum cooperation, to solve the multi-solution searching problems with an unknown number of solutions based on hybrid quantum-classical architecture. The algorithm is regarded to be a combination of the Partial Grover Search\cite{EG2004, EG2005} and the Depth-First Search\cite{DFS}. Besides this, we first proposed a method to deal with the difficulty of applying Grover Search to multi-solution problems.
    \subsection{Amplitude Interception}
    \par The general Grover search is not suitable for multi-solution searching problems\cite{ni2010-fs}. Some improvements were presented in \cite{BBHT1996, BCWZ}, however, there still exist the following flaws: \textbf{(1)} One will be possible to get repetitions, \textbf{(2)} For an unknown number of solutions the program has no criterion for terminating. In this section, analogous to amplitude amplification\cite{AA}, we introduce the concept of amplitude interception, which can solve both of the issues stated above.
    \par The general idea is to add a new gate to the head of each Grover's operations, where the gate $U_I$ is defined as follows
    \begin{equation}
        U_I\ket{x}\ket{y}=
        \begin{cases}
            \ket{x}\ket{y\oplus 1}\ \mathrm{for}\ x \in S\\
            \ket{x}\ket{y}\ \mathrm{for}\ x \notin S
        \end{cases}
    \end{equation}
    \par Where $S$ is the set of found solutions. In general, the ket $\ket{y}$ is initialized to $\ket{-}$ for phase kickback. Thus, applying $U_I$ gates has the effect of flipping the phase of founded elements. The new quantum circuit is shown in Figure 3.
    \begin{figure}[h]
        \begin{center}
            \begin{quantikz}
                \lstick{\ket{0}} & \qwbundle{n} & \gate{H} & \gate[wires=2][1cm]{U_I} \gategroup[2,steps=3,style={dashed, rounded corners, inner xsep=1pt, fill=blue!20}, background]{{Repeated several times}} & \gate[wires=2][1cm]{U_f} & \gate{U_s} & \meter{} & \cwbend{2} \\
                \lstick{\ket{-}} & \qw & \qw & \qw & \qw & \qw & \trash{\text{Trash}} \\
                \cw & \cw & \cw & \cw & \cw & \cw & \cw & \cw & \cw 
            \end{quantikz}
        \end{center}
        \vspace{-0.2cm}
        \caption{The modified structure of the Grover algorithm, with an additional gate added in front}
    \end{figure}
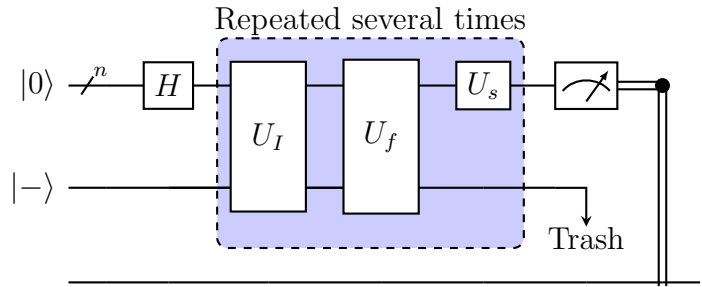
    \par In the circuit, $U_f$ and $U_s$ are still the oracle and diffuser\cite{Grover1996}. The combination $U_I\otimes U_f$ will flip and only flip the undiscovered solutions, so that found solutions will be regarded as the same as other trivial elements, and contributes amplitudes to the real solutions.
    \par Therefore, if we try to apply this amplitude interception version of Grover's search several times to find all solutions, we cross out the founded solutions before each trial and avoid repetitions. At the same time, with attachment to a classical computer, we can terminate the program immediately after getting wrong results at certain times, where the classical computer plays the role of register and validator.
    }
    {
    \subsection{Depth-First Grover Search}
    \par Suppose now we are doing searches on a database constructed by $N=2^n$ elements, with an unknown number of solutions. Taking the method above to do multi-solution search problems gives an average complexity of $\mathcal{O}(m\sqrt{N})$ and is expected to degenerate to $\mathcal{O}(N\sqrt{N})$ for the extreme case with all elements are solutions, which is even less efficient than a classical brute-force $\mathcal{O}(N)$ search. Therefore, the method aforementioned is still not optimal.
    \par In this section, we merged Partial Grover Search with Depth-First Search(DFS) and presented the Depth-First Grover Search(DFGS) with the help of amplitude interception to solve the multi-solution searching problems. The outline of DFGS is illustrated in Figure 4.
    
    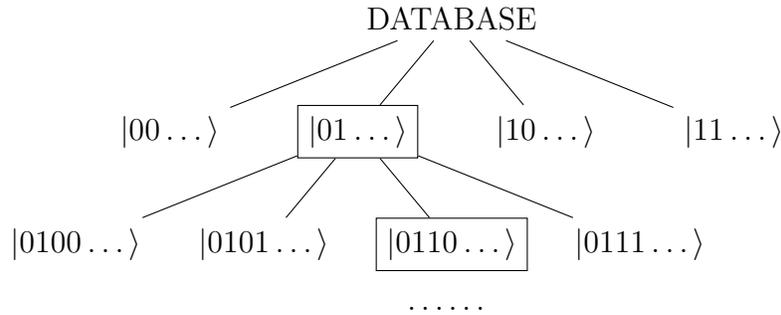
\begin{figure}[htbp!]
        \centering
        \begin{tikzpicture}[sibling distance = 2.5cm]
        \node {DATABASE}
        	child {node {$\ket{00\dots}$}}
        	child {node [draw]{$\ket{01\dots}$}
        	    child {node {$\ket{0100\dots}$}}
                child {node {$\ket{0101\dots}$}}
                child {node [draw]{$\ket{0110\dots}$}}
                child {node {$\ket{0111\dots}$}} 
        	}
        	child {node {$\ket{10\dots}$}}
        	child {node {$\ket{11\dots}$}};
        \end{tikzpicture}\\
        \ \ \ \ \ \ \ \ \ \ \dots\dots
            \caption{The steps of DFGS with $p=4$. $p=4$ means to split the database into $4$ pieces for each layer, or in other words, to determine $2$ bits of solution on each layer. Where the general idea is depth-first search, but the task of determining the next $2$ bits of the solution address is accomplished by a Partial Grover Search}
    \end{figure}
    \par To be specific, we assume one has the resources to do (partial) Grover searches with high precision for $p$ blocks. At each layer, we used Partial Grover Search to determine which intervals among $p$ equal-size splited sub-intervals contain solutions, and with DFS to recurse through every potential solution interval. Finally, for intervals with sizes less or equal to $p$, we adopt a normal Grover Search to find the exact address of the solution. 
    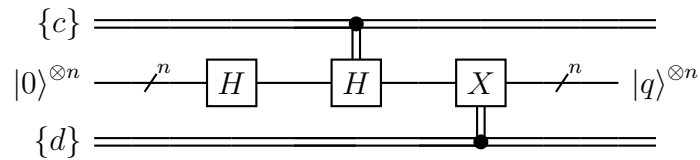
\begin{figure}[htbp!]
        \centering
        \begin{quantikz}
            \lstick{$\{c\}$} & \cw & \cw & \cw & \cw & \ \cw \cwbend{1} & \cw & \cw & \cw & \cw & \cw & \cw \\
            \lstick{$\ket{0}^{\otimes n}$} & \qw & \qwbundle{n} & \gate{H} & \qw & \gate{H} & \qw & \gate{X} & \qw & \qwbundle{n} & \qw \rstick{$\ket{q}^{\otimes n}$}\\
            \lstick{$\{d\}$} & \cw & \cw & \cw & \cw & \cw \cw & \cw & \cw \cwbend{-1} & \cw & \cw & \cw & \cw \\
        \end{quantikz}
        \caption{The modified $\mathtt{ENCODER}$ component, with two classical arrays. The classical array labeled $c$ plays the role of recording current bits and copying them to qubits when encoding. Another classical array labeled $d$ stores Bool values to cache if corresponding bits are determined. If one is determined, then the Hardmard gate is activated to convert the corresponding qubits from EPR superposition to $\ket{0}$ or $\ket{1}$. This $\mathtt{ENCODER}$ is applied to both the normal Grover Search and the partial one.}
    \end{figure}
    \par The amplitude interception is applied to Grover and Partial Grover searches. In the original paper on Partial Grover Search, the algorithm was designed to find the first $k$ bits of the solution index. In our algorithm, we expect the Partial Grover Search to find the \textit{next} $\ell=\log_2 p$ bits. Or in other words, the action we take is
    \begin{equation}
        |\underbrace{\dots\dots}_\text{Determined}\underbrace{???}_\text{next $\ell$}\dots\rangle\xrightarrow{\mathrm{Partial\ Grover\ Search}}|\underbrace{\dots\dots}_\text{Determined}\underbrace{abc}_\text{next $\ell$}\dots\rangle
    \end{equation}
    \par This can be implemented by modifying the $\mathtt{ENCODER}$ of the circuit, as shown in Figure 5. We put two auxiliary classical arrays for storing \textbf{(1)} the determined value of bits, \textbf{(2)} the state of the bits(Checked and Unchecked).  With this $\mathtt{ENCODER}$, the qubits will be initialized into a superposition of
    \begin{equation}
        \underbrace{\ket{\ \ \ \dots\dots\ \ \ }}_\text{Determined r bits} \otimes\ \frac{1}{\sqrt{2^{n-r}}}\bigoplus_{i=0}^{n-r}(\ket{0}+\ket{1})
    \end{equation}
    \par To clarify the process of $\mathtt{ENCODER}$, consider the example given in Figure 4. On the second layer, the first $4$ bits were decided, therefore two classical arrays $\{c\}$ and $\{d\}$ will be like $\{c\}=\{0,1,1,0,\dots\}$ and $\{d\}=\{1,1,1,1,\dots\}$, represents the first four bits are determined to be $0110$. Under these conditions, $\mathtt{ENCODER}$ will give
    \begin{equation}
        \ket{0110}\otimes\frac{1}{\sqrt{2^{n-4}}}\bigoplus_{i=0}^{n-4}(\ket{0}+\ket{1})
    \end{equation}
    
    \par The determined bits are fixed, and the Partial Grover Search will only find the next $k$ bits, which is our goal. So now we have everything we need to write down DFGS, as follows, implemented by recursions\cite{IntroAlg}.
    
    \begin{algorithm}[h]
    \renewcommand{\thealgorithm}{}
    \caption{Depth-First Grover Search(DFGS)}
        \begin{algorithmic}[1]
            \State $c\gets\{0\}_n$, $d\gets\{0\}_n$
            \Procedure{DFGS}{$r$, ans}
                \State $S\gets\emptyset$, $\ \mathrm{count}\gets0$
                \State Assign values to $c$ and $d$ for $\mathtt{ENCODER}$
                \If {$n-r\leq \ell$}
                    \State address $\gets$ Grover Search \Comment{Normal Grover search to find solutions}
                    \State ans $\gets$ ans $\ll \ell\ |$ address \Comment{Bitwise operations}
                    \State \Return $\{x\ |\ x\in \mathrm{ans},\ f(x)=1\}$
                \EndIf
                
                \While {count $<$ $p$}
                    \State address $\gets$ Partial Grover Search \Comment{Most possible sub-interval}
                    \State ans$'\gets$ ans $\ll \ell\ |$ address
                    \State Ret $\gets$ DFGS($r+\ell$, ans$'$)\Comment{Recursion}
                    \If {Ret = $\emptyset$} \Comment{The most possible interval gives no solution}
                        \State \textbf{break}
                    \EndIf
                    \State $S$ $\gets$ $S$ $\cup$ Ret, count $\gets$ count + 1
                \EndWhile
                \State Recover elements in $c$ and $d$
                \State \Return $S$
            \EndProcedure
            \State $\textbf{end procedure}$
        \end{algorithmic}
    \end{algorithm}
    \par Where bitwise operations were used for simplicity. In line 6, we expect the performance of Grover Search will find all possible solutions in the given interval, so the resultant addresses and therefore $\mathtt{ans}$ are sets. In line 8 we use the classical part to do further verification and abandon all incorrect solutions in the answer set\ (for $f(x)\neq 1$ is an incorrect solution).
    \par With recursion, we anticipate DFGS will have the ability to find all solutions. After looking up one interval, the algorithm will backtrack to the previous layer and continue recursing into the next possible interval (line 12) until the most-possible intervals give no solutions (line 13).
    }

    \subsection{Analysis on Performance}
    {
    \subsubsection{Time complexity}
    \par Some rough analyses tell the complexity of the DFGS algorithm will have an average complexity of $\mathcal{O}(m\sqrt{N})$, and $\mathcal{O}(\sqrt{p}N)$ for every element is a solution.
    \par For each recursion, the program performs either partial Grover searches or normal Grover searches (for intervals with a size less or equal to $p$). According to \cite{EG2004}, the total queries required for one partial Grover Search is $\sqrt{L}-\sqrt{L/p}$, where $L$ refers the length of searching interval for recursion depth $k$, the number of queries is $\sqrt{N/p^k}-\sqrt{N/p^{k+1}}$. The maximum depth of recursion, $\lambda$, can be solved out by
    \begin{equation}
        \frac{N}{p^\lambda}=p\ \ \rightarrow\ \ \lambda=\frac{\log N}{\log p}-1
    \end{equation}
    \par We suppose at depth $k$ there are $\alpha_k$ partial searches performed $(1\leq\alpha_k\leq p)$, then the complexity of DFGS will be
    \begin{equation}
        \sum_{k=0}^{\lambda-1} \alpha_k\left(\sqrt{\frac{N}{p^k}}-\sqrt{\frac{N}{p^{k+1}}}\right)+\alpha_{k+1}\sqrt{\frac{N}{p^{\lambda}}}
    \end{equation}
    \par $\alpha_k$ represents the expected number of Partial Grover Searches performed on layer $k$. For the case with $\alpha_k\equiv 1$, corresponds to a situation with only one solution, DFGS has a strict complexity of $\mathcal{O}(\sqrt{N})$, and for the case with all elements are solutions, $\alpha_k\equiv p^k$, DFGS has an asymptotic complexity of $\mathcal{O}(\sqrt{p}N)$. 
    \par To figure out the average complexity, $\alpha_k$ should be numerically equal to the expected number of intervals that involve solutions. We define
    \begin{equation*}
        x_i=
        \begin{cases}
            1\ \mathrm{for\ interval\ with\ no\  solutions}\\
            0\ \mathrm{for\ interval\ with\ any\ number\  of\ solutions}
        \end{cases}
    \end{equation*}   
    \par Where, the index $i$ refers to the $i$th block on current layer $k$\ (where there are $p^k$ blocks in total). Therefore $X=\sum x_i$ is the number of blocks with no solutions. We have the relationship
    \begin{equation}
        E(X)=E\left(\sum_{i=0}^{p^k}x_i\right)=\sum_{i=0}^{p^k}E(x_i)
    \end{equation}
    \par $E(x_i)$ is relatively easier to figure out, the result is $E(x_i)=(1-p^{-k})^m$, and so $E(X)=p^k(1-p^{-k})^m$, then the expectation number of intervals with solutions, $\alpha_k=p^k(1-(1-p^{-k})^m)$. So the summation can be written as
    \begin{equation}
        \sum_{k=0}^{\lambda-1} p^k(1-(1-p^{-k})^m)\left(\sqrt{\frac{N}{p^k}}-\sqrt{\frac{N}{p^{k+1}}}\right)+m\sqrt{\frac{N}{p^{\lambda}}}
    \end{equation}
    \par Apply Bernoulli's inequality
$(1+x)^r\geq1+rx$, we have $(1-(1-p^{-k})^m)\leq mp^{-k}$
    \begin{equation}
        \begin{split}
            \mathrm{Equation}\ (8)&\leq\sum_{k=0}^{\lambda-1} m\left(\sqrt{\frac{N}{p^k}}-\sqrt{\frac{N}{p^{k+1}}}\right)+m\sqrt{\frac{N}{p^{\lambda}}}\\
        \end{split}
    \end{equation}
    \par After finishing this summation, we conclude that DFGS has an average complexity of $\mathcal{O}(m\sqrt{N})$. In comparison to a normal Grover Search, the Depth-First Grover Search(DFGS) attains the same average complexity but a better performance for problems with very densely-distributed solutions\ (it will not worsen to $\mathcal{O}(N\sqrt{N})$). Or in other words, DFGS is more stable for unstructured multi-solution problems.
    \par The remaining question is the determination value of $p$. Qualitative analysis will suggest that $2$ is a reasonable choice of $p$, the smallest integer greater than $1$\ (which corresponds to a classical brute-force search, with complexity $\mathcal{O}(N\sqrt{N})$), or in other words to find a single bit on each layer. The selection of $p=2$ will work outstandingly for an evenly-distributed database\ (including the case where all elements are solution). On the other hand, for the remarkable case of a densely-distributed database, a larger searching block\ (i.e., a greater $p$) is expected to be more efficient in determining intervals of solutions and catching all of them in one draft.
    \par In general, the value of $p$ will closely rely on the predicted configuration of the database and the quantum resources available, so in this paper, we will not give a specific upper bound for the value $p$ but keep it as a manually determined parameter.
    }
    {
    \subsubsection{Correctness}
    \par Complexity aside, the algorithm has an unavoidable possibility of not managing to figure out all solutions. In an extreme case, the program may go astray on the first layer and directly cause the whole algorithm to end up with no solutions. One method to eliminate this risk is to change the criteria of terminating, to make the program terminate until it finds incorrect solutions(or no solutions) for $q_k$ times on layer $k$. Since for deeper layers, the number of elements is decreasing, $q_k$ is expected to be inversely proportional to $k$, as follows
    $$q_k=\left \lceil{\frac{\nu}{k+1}}\right \rceil,\ \ \ \ \ \  \alpha'_k=\alpha_k+q_k$$
    \par Again, $\nu$ is still a manually determined value but with an upper bound $p$. The value of it depends  on the goal accuracy and the number of elements. The corresponding average complexity for DFGS with new $q_k$ becomes $\mathcal{O}((m+\nu)\sqrt{N})$. Since the coefficients of summation have the following relationship
    $$\alpha_k+\left \lceil{\frac{\nu}{k+1}}\right \rceil\leq\alpha_k+\nu$$
    \par After implementing this new criterion, the algorithm will then have greater opportunities to sort out all solutions.
    }
}

\section{Conclusion}
    \par In this paper, we proposed a quantum advantage algorithm DFGS, with quantum-classical computer hybridizing to do search problems for an unknown number of solutions. With the help of amplitude interception, our novel algorithm has the ability to figure out all solutions and attains the same efficiency as the normal Grover search but a more robust and stable performance for extreme cases.
    \par On the implementation level, the main body of DFGS takes an equal number of Grover's diffusers and measurements. An extra $\mathtt{ENCODER}$ is required for intercommunication between the classical and quantum parts. For the $\mathtt{ENCODER}$, besides Hardmard gates, several gates controlled by the classical part need to be arranged. Moreover, the new gate $U_I$ added for amplitude interception can be absorbed by adjusting the $U_f$ gate.
    \par Based on the current trends of quantum computation, more and more quantum algorithms will rely on the hybrid quantum-classical computer. On the other hand, when designing classical algorithms, one can also consider embedding quantum algorithms for optimal performance. For example, a heap structure is difficult on determining the exact index of a specific element \cite{Fredman_1987}. Applying a quantum search with complexity $\mathcal{O}(\sqrt{N})$ may be a solution.
    \par In conclusion, Well-designed hybrid quantum-classical based algorithms or data structures can be \emph{universal}, which means capable of handling any circumstances and requirements. The Depth-first Grover Search algorithm is illustrated to be \emph{universal} since its capability for any configuration of search problems. Besides it, tons of hybrid algorithms with better performance are waiting to be discovered.

\printbibliography

\end{sloppypar}
\end{document}